\documentclass[aps,prb,preprint,groupedaddress]{revtex4-2}

\usepackage{float}
\usepackage{graphicx}
\usepackage{url}
\usepackage{caption}
\usepackage{adjustbox}
\usepackage{amsmath}
\usepackage{physics} 
\usepackage{xcolor}

\newcommand{\angstrom}{\mbox{\normalfont\AA}}

\begin{document}

%\preprint{}

\title{Suppression of Peierls-like, nesting-based instabilities in solids}


\author{Nassim Derriche}
\author{Ilya Elfimov}
\author{George Sawatzky}
%\thanks{}
\affiliation{Quantum Matter Institute, University of British Columbia}


%\date{\today}

\begin{abstract}
Understanding instabilities is of vast importance in material science. One such example is the Peierls instability of one-dimensional metals with a single band. The question is why H and Li chains behave so differently in spite of their similar electronic structure. We demonstrate that this is due to the suppression in the susceptibility for Li because of the interatomic, inter-orbital s-p mixing of the band crossing the Fermi energy. The influence of this on the structure factor-like term in response functions is critical and depends on the wavefunction’s details also in higher dimensions.
\end{abstract}

%\keywords{}


\maketitle

%\section{Introduction}

The susceptibility of a system is a central theme in all of condensed matter physics since it is what we need to know and understand in order to describe the response of a system to external perturbations. This includes temperature, pressure, magnetic or electric fields, electron density variations due to doping and chemical substitutions as well as the influence of various interactions such as the electron-boson interactions (be it electron-hole excitations, phonons, magnons, plasmons, etc.). In addition, the momentum and frequency dependence of the susceptibility provides information regarding possible phase transitions such as insulator-metal transitions as well as charge and spin density wave formations, magnetic ordering of various kinds and superconductivity. Presently, the approximation mostly used for the static charge susceptibility is the Lindhard function as derived for a free electron gas involving only the band structure and ground state occupation numbers. However, the generalized form involves a multiplicative structure factor-like term which has a substantial momentum dependence if the Bloch wavefunctions involve strongly localized atomic wavefunctions and strongly corrugated low energy scale charge densities. We demonstrate the extreme importance of the detailed nature of the wavefunctions represented by the structure factor-like term that multiplies the purely band structure-based Lindhard susceptibility. We show that, especially for atomic-based bands of opposite parity with a strong k-dependent mixing because of interatomic hybridization, which is generally the case for materials of interest, the response of a system can change dramatically from that determined using only the band dispersion. This is especially important in the case of so-called topological materials and their surfaces but also for all systems in which mixing between opposite-parity orbitals such as s and p, d and p, or d and f atomic orbitals is involved in the low energy scale wavefunctions.

In free electron gas models at zero temperature, the charge susceptibility diverges in 1D, the first derivative diverges in 2D and a kink occurs in 3D at a wave vector equal to $2k_F$. This leads to a potential instability to an external potential at a wave vector of $2k_F$, giving rise to phase transitions such as the predicted Peierls transition for a half-filled 1D band \cite{peierls_original}, charge density waves prominent in 2D and the Friedel charge density oscillations about a charged impurity in a 3D metal. In more realistic systems, we have to deal with a lattice of atoms and a strongly corrugated potential landscape. Furthermore, the Fermi surfaces in 2D and 3D materials differ strongly from the free electron circular or spherical Fermi surfaces and, in cases where large parts of the Fermi surface can be connected with the same q vector (nesting), the charge susceptibility in a single band model in tight binding theory would diverge as in the 1D case. For the high-$T_c$ superconductors such as the Fe pnictides \cite{pnictide_nesting}, this nesting property has been suggested as the origin of the superconductivity itself. In 2 dimensional systems such as the HTC cuprates \cite{cuprate_cdw} and two dimensional dichalcogenides (TMD’s) \cite{di_nesting}, Fermi surface nesting can lead to charge density waves as highlighted in recent scientific activity. Also, recent studies of the 3d transition metal perovskite structure oxides as in the rare-earth nickelates have demonstrated structural and phase transitions suggested to be related to nesting conditions \cite{Balents_nesting}. Some of these are strongly related to Peierls-like transitions involving strong atomic movement effectively as a result of electron-phonon coupling. There is a significant focus on charge density waves in higher-dimensional materials in contemporary literature \cite{kohn_anomaly}. To a lesser extent, ab initio exploration of 1D systems in order to study Peierls transitions, especially from a phononic perspective, is also present for 1D materials such as HF chains \cite{hf_chain} in the literature.

Recently, several authors using density functional theory (DFT) methods have questioned the theoretical validity of Peierls' original conclusions and the causes of charge density waves in some materials \cite{c_chain_inadequate}. Notably, Johannes and Mazin argued in 2008 against a naive application of Peierls theory to real CDW materials. They analyzed cases where charge density waves exist with wavevectors at which Fermi surface nesting is weak or non-existent. They also showed that, using DFT, a linear chain of equally spaced sodium atoms with one valence electron per atom is not susceptible to a transition to a stable 1D dimerized phase \cite{mazin_na_2008}. The phase diagram of hydrogen chains has been studied through various ab initio and computational methods and in all cases clearly shows a dimerized ground state. However, the root cause of the absence of dimerization for sodium has not been explored \cite{h_chain_computational_methods,h_chain_multiple_methods,h_chain_monte_carlo}. DFT calculations result in dimerization for 1D H but not for 1D Li or Na \cite{dft_na}. In this paper, we demonstrate that this results from a very basic difference in the atomic electronic structure and the interatomic, inter-orbital hybridization involving the mixing of even and odd parity atomic wavefunctions. Such mixing is present in many systems, excluding H and 1D organic molecular systems in which the valence electrons are in orbitals with intermolecular $\pi$ bonds such that even-odd parity intermolecular hybridization is symmetry forbidden, like in carbon chains \cite{ssh,c_strain,c_chain_true}. Materials with s-p mixing like the alkali metals clearly incorporate such interactions, but systems that involve higher angular momentum orbitals also have the potential to involve mixing with similar symmetry.

\begin{figure*}
\includegraphics{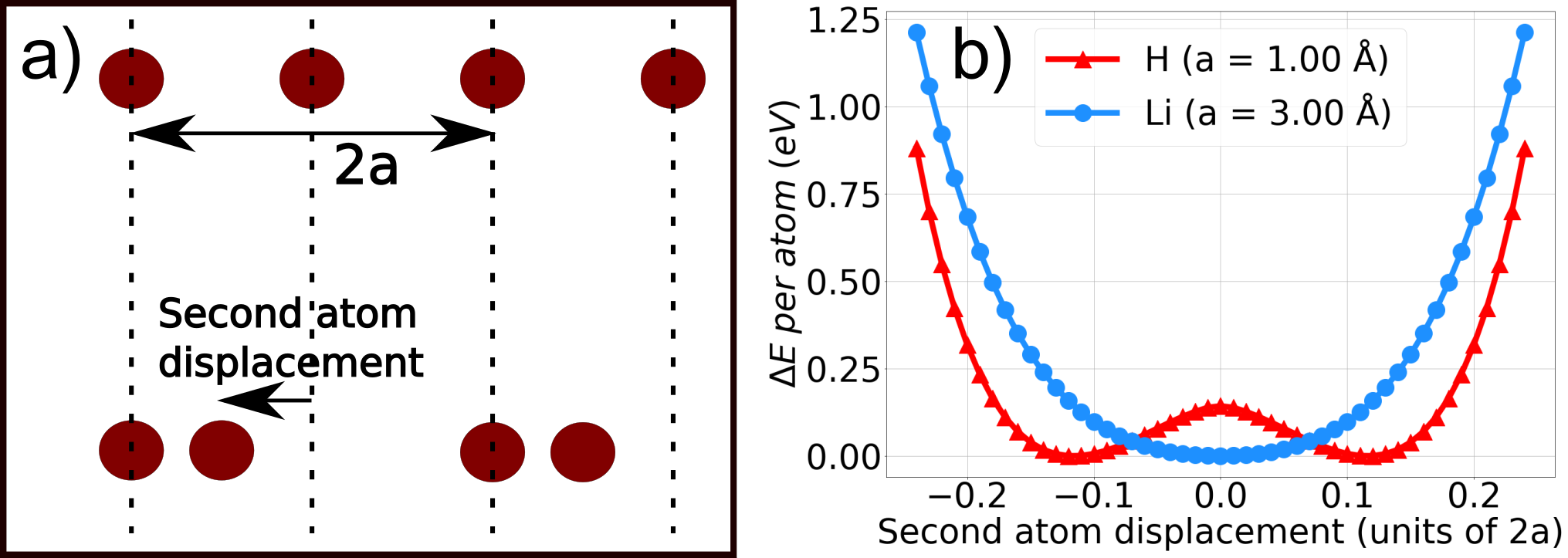}
\caption{a) Diagram of an undisturbed 1D configuration (upper chain) and a configuration with a non-zero dimerization (lower chain). b) Total DFT energy change as a function distance between neighboring atoms. The energy difference between the ground state and the undisturbed state of H is 143 meV.}\label{fig:dimer_energy}
\end{figure*}

We study the changes that occur in the charge susceptibility when moving from H to the alkali metals. First, we follow the calculations of Johannes and Mazin, but now for H and Li chains through DFT. The local density approximation was used for the exchange-correlation functional \cite{perdew}, and the basis set was atomic orbital based (FPLO) \cite{fplo}. Equilibrium lattice parameters were determined through relaxation of one-dimensional chains of equally-spaced atoms separated by 40 $\angstrom$, which totally suppresses inter-chain interactions, resulting in
1.00 $\angstrom$ for H (compared to 1.41 $\angstrom$ in a simple cubic structure \cite{Note1}), and 3.00 $\angstrom$ for Li compared to 3.04 $\angstrom$ for the experimental nearest-neighbor distance in BCC Li metal. In Figure \ref{fig:dimer_energy}, we show the total energy change as a function of the degree of dimerization schematized in Figure \ref{fig:dimer_energy} a). The dimerized H chains have a lower energy by about 143 meV or $k_BT$=1659.5 K compared to the uniform chain, while the Li chain has its lowest energy at the uniform configuration. This raises the question as to the origin of this dramatic difference.

While the electronic configuration of these two elements is similar in the sense that they both have only one valence electron in an s symmetry orbital as atoms, the key difference is that the principal quantum number of 1 can only house a 1s orbital, while an n=2 level can house s and p orbitals which are relatively closely spaced in energy (a difference of 2.3 eV in Li due to relativistic effects). In atoms, this difference is not really of basic importance for the ground state but in a solid one must deal with interatomic inter-orbital hybridization. The s and p bands in 1D have opposite dispersion and if their bandwidths are large enough, they would cross were it not for the also strong interatomic s-p hybridization which lifts the degeneracy at the crossing point and opens a substantial gap leading to a well-separated single lowest energy band whose bandwidth has quite dramatically decreased from that of a 2s only dispersion. However, since the s-p interatomic hybridization vanishes at $k = 0$ and $k = \pi$, the character of the lowest band is purely s at $k = 0$ and purely p at $\pi$.

\begin{figure*}
\includegraphics{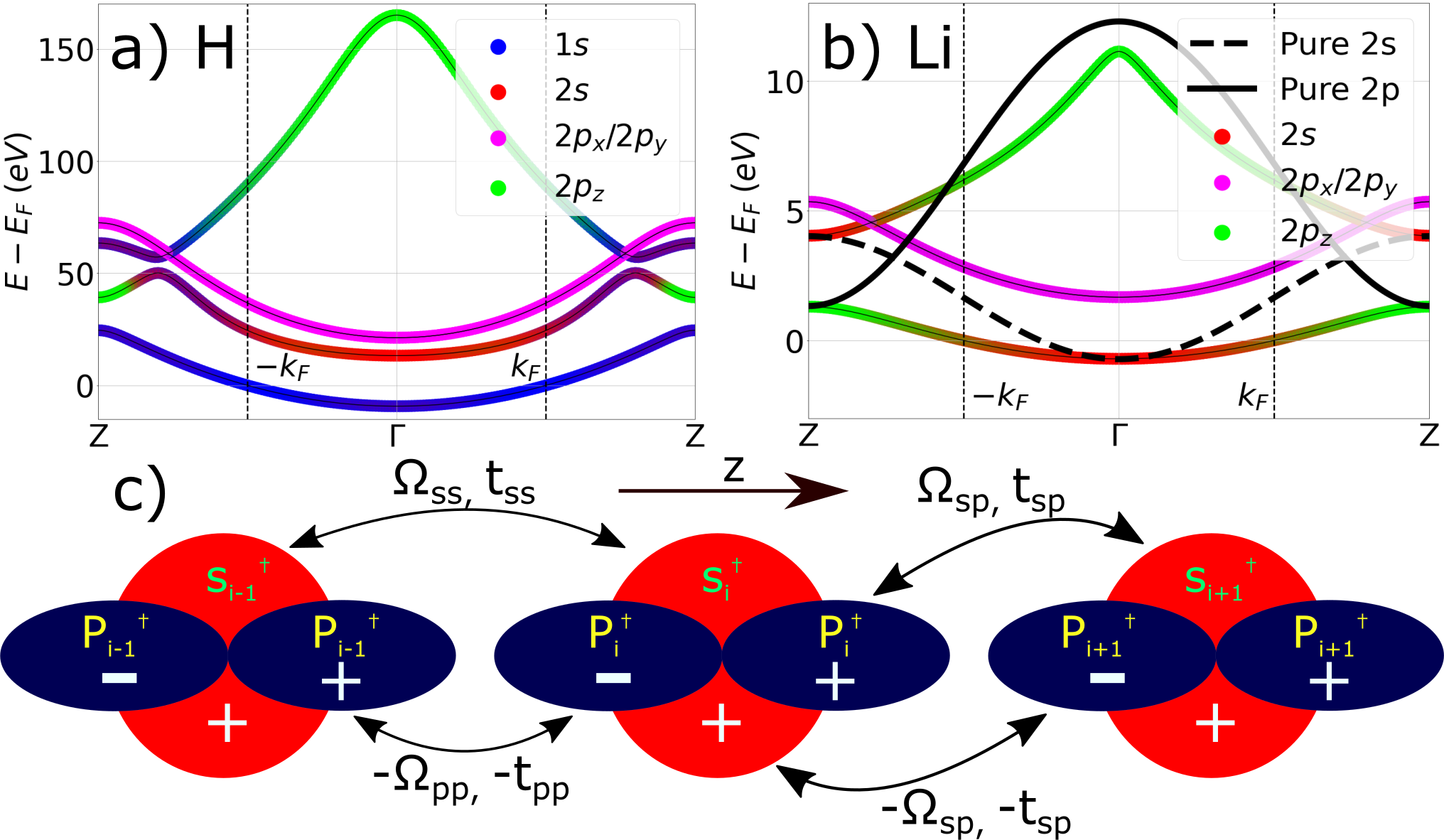}
\caption{DFT orbital-projected band structures for undisturbed chains of a) hydrogen, b) lithium. The dashed black lines corresponds to tight binding, nearest neighbor purely 2s and 2p dispersions.  c) Schematic of the real space 1D lithium chain model considered for the tight binding Hamiltonian in Equation \ref{eq:hamiltonian}. 2s and 2p$_z$ orbitals are shown (including + and - signs indicating phases) with their associated creation operators, along with the hopping parameters and overlap integrals taken into account. The hopping parameters used are discussed in the text.}\label{fig:bands}
\end{figure*}

All of the above effects are displayed in the DFT calculations displayed in Figure \ref{fig:bands}. There is only one band crossing the Fermi energy for Li as for H, but the H band is purely of s character and separate from higher energy bands by 17 eV while the Li band has a strongly momentum-dependent character which changes from s to p going from $\Gamma$ to the zone boundary. As we will demonstrate below, it is the interorbital hybridization of \textit{gerade} (2s) and \textit{ungerade} (2p) orbitals that causes the suppression of the divergence in the susceptibility in a chain of Li atoms.

The susceptibility involves not only the scattering of electrons from occupied to unoccupied states, but also the spatial part of the wavefunctions. In the free electron approximation, the charge density for each k state is simply a constant independent of k. In a solid with atoms, the density varies strongly within the unit cell. This difference is often ignored \cite{plane_wave}. In a solid with translational symmetry, the wavefunctions are known to be Bloch waves which, in the limit of large atomic spacings, can be decomposed into linear combinations of
atomic orbitals (LCAO). The electronic states in a single band can thus be expressed as:

\begin{equation}\label{eq:wavefunction}
\Psi_{\vec{k}}(\vec{r}) = \frac{1}{\sqrt{N}}\sum_{\mu,l} e^{i\vec{k} \cdot \vec{R}_l} \alpha_\mu (\vec{k}) \phi_\mu(\vec{r}-\vec{R}_l),
\end{equation}

\noindent where $\phi_\mu(\vec{r})$ is the real space form of the atomic orbital $\mu$, $R_l$ is the distance of the l'th atom from the origin along the chain, N is the number of atoms and $\alpha_\mu(\vec{k})$ is the normalized k-dependent coefficient of the multi-atomic wave function ($\mu$ is 2s or 2p in our case). 

The static one-dimensional, random phase approximation single band susceptibility \cite{textbook_susc,full_susc,susc_scat,sham_original,sham_1,sham_2}, can be written using a LCAO basis as:

\begin{equation}\label{eq:susceptibility}
\chi(q) = \sum_{k} \frac{f_k - f_{k+q}}{\epsilon_k - \epsilon_{k+q} + i\eta} n(k,q) = \sum_{k} \chi_0(k,q) n(k,q)
\end{equation}

\begin{eqnarray}\label{eq:scat_fact}
n(k,q) = \sum_{\mu,\nu,l, \mu', \nu', l'} e^{i(k+q)(R_l - R_{l'})} \alpha^\ast_\nu(k)\alpha_\mu(k+q) \alpha_{\mu'}(k+q)\alpha_{\nu'}(k) \\ \times \mel{\phi_{\nu}(\vec{r})}{e^{-i\vec{q} \cdot \vec{r}}}{\phi_{\mu}(\vec{r} - \vec{R}_l)} \mel{\phi_{\mu'}(\vec{r} - \vec{R}_l')}{e^{i\vec{q} \cdot \vec{r}}}{\phi_{\nu'}(\vec{r})} \nonumber
\end{eqnarray}

\noindent where $\epsilon_k$ is the energy of the band at k, $f_k$ is the Fermi-Dirac distrubtion at k, $i\eta$ is a small imaginary part representing a finite state lifetime and the k sum is over the first Brillouin zone. We see here that the n(k,q) factor directly depends on the band orbital character. When modeling the charge carriers as plane waves, Equation \ref{eq:susceptibility} leads to the standard Lindhard function in 1D which is characterized by a divergence at $q=2k_f$ \cite{peierls_figure,don}. On the other hand, taking the approximation that the atomic wave functions are real space delta functions centered on lattice sites conserves the $\alpha(k)$ coefficients, but the contributions from the Fourier integrals in Equation \ref{eq:scat_fact} become trivial \cite{only_coeffs}.

Because the DFT codes are concentrated on the electron density and the total energy, they do not provide the details of the wavefunctions needed to determine n(k,q). Therefore, we turn to a tight binding fit to the band structure using an atomic basis set for each atom, which we limit to the include only the valence s and p orbitals or the orbitals that are the main contributors to the valence electron band structure. This also directly includes the s-p interatomic hybridization as discussed above. We also take the non-orthogonality of the 2s and 2p states at different sites into account by including their nearest neighbor overlap integrals \cite{non_orthogonal}. The Hamiltonian in a (s,p) basis in the momentum representation is:

\begin{equation}\label{eq:hamiltonian}
H =  \sum_k \mqty(2(\Omega_{ss} \epsilon_k-t_{ss}) \cos{(ka)} & 2i(\Omega_{sp} \epsilon_k - t_{sp}) \sin{(ka)} \\ -2i(\Omega_{sp} \epsilon_k - t_{sp}) \sin{(ka)} & -2(\Omega_{pp} \epsilon_k - t_{pp}) \cos{(ka)} + \Delta),
\end{equation}

\noindent where $\Delta$ is the on-site energy difference between the p and s orbitals, $\epsilon_k$ is the energy dispersion (eigenvalues solved self-consistently), $t_{ss}$, $t_{pp}$ and $t_{sp}$ are nearest neighbor hopping parameters between neighboring s-s, p-p and s-p orbitals respectively and $\Omega_{ss}$, $\Omega_{pp}$ and $\Omega_{sp}$ are the nearest neighbor overlap integrals between the orbitals denoted in the subscripts. All these parameters are taken to be always positive; the phase relationships are encoded in the signs present in the Hamiltonian above. The imaginary off-diagonal terms encode the k-dependence of the s-p hybridization which, as symmetry dictates, is zero at $\frac{\pi}{a}$ and $\Gamma$, and is maximal at $\frac{\pi}{2a}$. Although the contributions to n(k,q) from the spatially-extended atomic orbitals themselves are significant, the most important contribution comes from these off-diagonal terms and the fact that the s and p orbitals are \textit{gerade} and \textit{ungerade} respectively, and the onsite mixing vanishes while the inter-site mixing changes sign when moving to the right or the left. This is critical to our demonstration.

A visualization of the orbitals and of the relevant parameters is provided in Figure \ref{fig:bands} c). Diagonalization of the Hamiltonian in Equation \ref{eq:hamiltonian} results in the energy eigenvalues $\epsilon(k)$ and eigenstates in momentum space, giving us a functional form for $\alpha_{2s}(k)$ and $\alpha_{2p}(k)$ in Equation \ref{eq:wavefunction}. 

Numerical values for the hopping parameters and the onsite energy difference were obtained by fitting the energy eigenvalues to our DFT results, which lead to $\Delta$ = 5.163 eV, $t_{ss}$ = 1.185 eV, $t_{pp}$ = 2.743 eV and $t_{sp}$ = 1.558 eV. To compute the overlap integrals as well as the matrix elements in Equation \ref{eq:scat_fact}, we used the analytical form of the hydrogenic orbital wavefunctions in 3D, which were adapted for the lithium case with an effective nuclear charge of $Z = +1.26e$ \cite{slater_rule}. The overlap integrals, using the DFT relaxed lattice constant, are $\Omega_{ss}$ = 0.553, $\Omega_{pp}$ = 0.276 and $\Omega_{sp}$ = 0.390. 

\begin{figure*}
\includegraphics{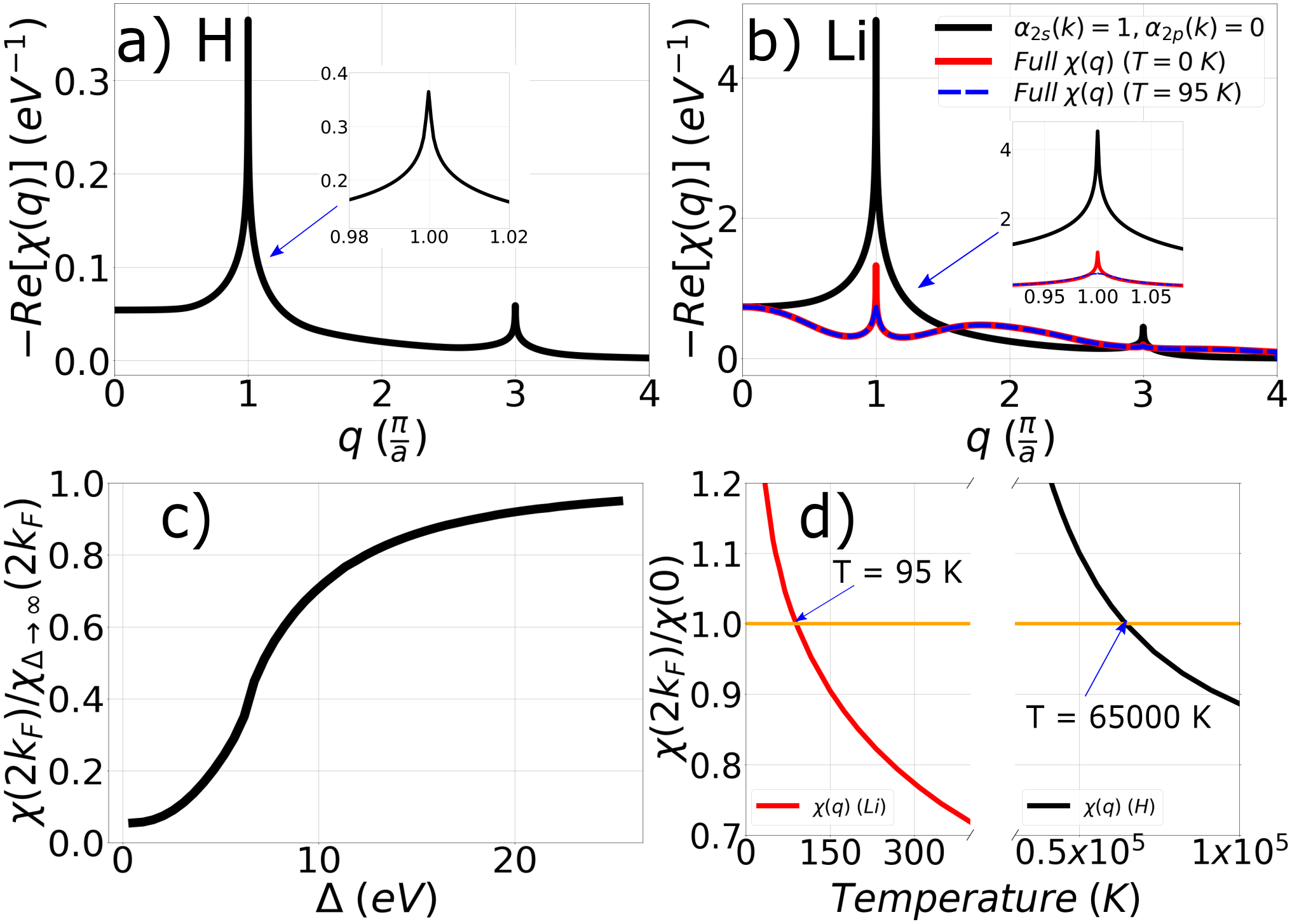}
\caption{Charge susceptibility of an undisturbed 1D chain of a) hydrogen and b) lithium using DFT energy data and eigenfunctions obtained from a tight binding fit to the DFT band structure. The effect including s-p hybridization (in red at zero temperature, and in blue at T=95 K) compared to enforcing a purely 2s state (in black at T=0) is shown for Li. The insets show the peaks at q = 2$k_F$. c) Ratio of the peak height of the full lithium charge susceptibility and of the susceptibility calculated with a very large charge transfer energy $\Delta$ as a function of $\Delta$. d) Ratio of the charge susceptibility peak height and its value at $q=0$ as a function of temperature for both H and Li. The orange lines correspond to the temperatures at which $\chi(2k_F)=\chi(0)$.}\label{fig:chi_Combined}
\end{figure*}

We proceed to show that the inclusion of the $n(k,q)$ factor exhibits the large difference in the susceptibility between H and Li chains. We follow Equation \ref{eq:susceptibility}, in which we only consider terms in the infinite sum appearing in Equation \ref{eq:scat_fact} which include on-site and nearest neighbour integrals ($\abs{l-l'}$ = 0 or 1) since they are the main numerical contributions. We set $\eta = 1 \; \mu eV$ for both H and Li, allowing us to numerically compare the apparent strength of divergences by making them finite width integrated Lorentzians even at zero temperature. As expected, we do obtain a divergence at $q=2k_F$ for H as shown in Figure \ref{fig:chi_Combined} a). On the other hand, for Li in Figure \ref{fig:chi_Combined} b), there is a strong suppression of the peak at $q=2k_F$ compared to the $\alpha_{2s}(k)=1$ case. Note that the value at $q=0$ is equal to the density of states at the Fermi level. Ignoring the orbital wavefunction-dependent matrix elements causes a smaller suppression of the peak and the expected decrease of the overall susceptibility going
as $\frac{1}{q^2}$ is not captured.

\begin{figure}
\includegraphics{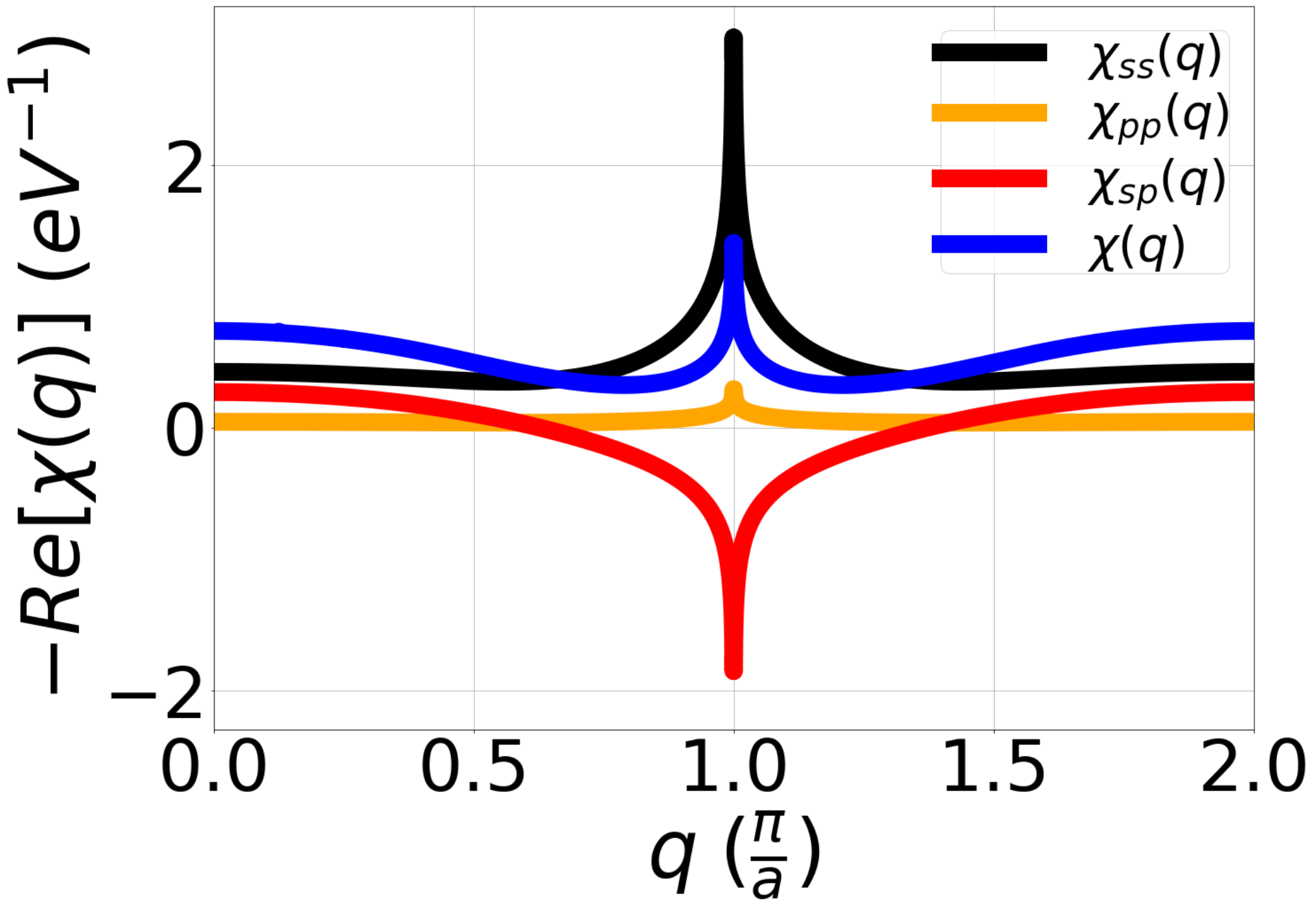}
\caption{Orbital contributions to the lithium one-dimensional chain susceptibility with respect to q as outlined in Equation \ref{eq:different_terms}. The susceptibility $\chi(q)$ (in blue) is the sum of the three other individual plotted terms.}\label{fig:terms_comparison}
\end{figure}

The tight binding model described by Equation \ref{eq:hamiltonian} is powerful enough to mathematically demonstrate the influence of the terms of different orbital dependence entering in $\chi(q)$. Taking the $\phi(\vec{r})$ terms as delta functions centered on atomic sites in Equation \ref{eq:scat_fact}, and thus considering only onsite terms ($\abs{l-l'}=0$), we can write the susceptibility as:

\begin{eqnarray}\label{eq:different_terms}
\chi(q) = \sum_{k} \chi_0 (k,q) \Big( \abs{\alpha_{2s}(k)}^2\abs{\alpha_{2s}(k+q)}^2 + \abs{\alpha_{2p}(k)}^2\abs{\alpha_{2p}(k+q)}^2 \\ + 2\alpha_{2s}^*(k)\alpha_{2s}(k+q)\alpha_{2p}^*(k+q)\alpha_{2p}(k) \Big) = \chi_{ss}(q) + \chi_{pp}(q) + \chi_{sp}(q) \nonumber,
\end{eqnarray} 

where $\chi(q)$ is decomposed into terms depending only on the eigenstate coefficients associated with the orbitals indicated by the associated subscripts. The analytical form of $\alpha_{2s}(k)$ and $\alpha_{2p}(k)$ obtained through diagonalization are lengthy functions of the Hamiltonian parameters, but they can be written in such a way that $\alpha_{2s}(k)$ is purely real and positive for all k, while $\alpha_{2p}(k)$ is purely imaginary for any k but is positive for $k<0$ and negative for $k>0$ (as is shown in the Appendix). Consequently, following Equation \ref{eq:different_terms}, we can see that $\chi_{ss}(q)$ and $\chi_{pp}(q)$ have the same sign for all q, but that $\chi_{sp}(q)$ can be positive or negative; $\alpha_{2p}(k+q)$ has the opposite sign of $\alpha_{2p}(k)$ at values of q for which k+q lands on the other half of the Brillouin zone compared to k. The number of terms in the k sum for which this holds increases as q approaches $2k_F$, which results in a dimming of the susceptibility peak in the neighborhood of that value. These different contributions are shown clearly in Figure \ref{fig:terms_comparison}. This mathematical behavior is a direct consequence of the \textit{gerade}-\textit{ungerade} nature of the 2s-2p hybridization.

In addition, the full susceptibility peak height is shown as a function of the energy difference between the 2s and 2p orbitals in Figure \ref{fig:chi_Combined} c). We see that the reduction in hybridization with increasing $\Delta$ reduces the $\chi(q)$ peak monotonically, approaching the pure 2s result.
To get an impression concerning the phase stability of the uniform phase of Li relative to that of H, we show in Figure \ref{fig:chi_Combined} d) the ratio of the susceptibility at $q=2k_F$ and at $q=0$. The fully hybridized lithium system is more susceptible to long range (small q) charge density variations than to a doubling of the unit cell at temperatures above about 95 K (an energy of 8 meV when using the electronic temperature $k_BT$); $\chi(q)$ at that temperature is plotted in Figure \ref{fig:chi_Combined} b). However, the $q=2k_F$ peak of the H chain remains the global maximum even at very high temperatures. We note that the temperature where $\chi(2k_F) = \chi(0)$ should not be taken as the expected, definite transition temperature since we also have to take into account the changes that would occur in the other degrees of freedom such as the lattice energy represented by phonons in order to determine it. It is however a strong indication that the phase transition, if it occurred at all, would be at a much lower temperature than  that suggested from the susceptibility not including the inter-orbital hybridization and overlap integrals. 

We thus conclude that a direct investigation of the Fermi surface of a material to look for nesting q-vectors is not enough even for qualitative predictions of instabilities in systems for which the wavefunctions are strongly non-free electron-like. Although we have demonstrated this for the case of 1D and emphasized the even stronger influence of interatomic, inter-orbital mixing terms involving \textit{gerade} and \textit{ungerade} symmetry wavefunctions, this will be equally important in 2 and 3 dimensions, especially when closely spaced \textit{gerade} and \textit{ungerade} states are involved in the low energy states. This will not always suppress the susceptibility peaks, but very likely could cause peaks to occur in other regions in momentum space. The importance of including n(k,q) could apply especially to topological materials, but also to transition metals and rare earth compounds in which there is strong interatomic hybridization of \textit{gerade} and \textit{ungerade} states, including for example the high-$T_c$ superconductors and topological materials.

\begin{acknowledgments}

We thank Kateryna Foyevtsova and Oliver Dicks for fruitful discussions regarding DFT and susceptibility theory. This research was undertaken thanks in part to funding from the Max Planck-UBC-UTokyo Center for Quantum Materials and the Canada First Research Excellence Fund, Quantum Materials and Future Technologies Program, as well as by the Natural Sciences and Engineering Research Council (NSERC) for Canada.

\end{acknowledgments}

\appendix*
\section{Tight Binding Model Eigenstates}

Diagonalizing the tight binding Hamiltonian shown in Equation 4 leads to the eigesnstates of our system. The lower energy eigenstate, which represent the bonding interaction between the 2s and 2p Li orbitals, is shown as the band crossing the Fermi energy in Figure 2 b). The form of the coefficients $\alpha_{2s}(k)$ and $\alpha_{2p}(k)$ in Equation 1 for that eigenstate, when setting the overlap integral values to zero for clarity, are as follows:

\begin{align}\label{eq:2s}
\alpha_{2s}(k) = \frac{4 t_{sp} \abs{\sin(ak)}} {\sqrt{16 t_{sp}^2 \sin^2(ak) + \gamma^2(k)}},
\end{align}

\begin{align}\label{eq:2p}
\alpha_{2p}(k) = \frac{i\gamma(k) \abs{\sin(ak)}} {\sin(ak)\sqrt{16 t_{sp}^2 \sin^2(ak) + \gamma^2(k)}},
\end{align}

\begin{align}\label{eq:gamma}
\gamma(k) =  2(t_{ss} + t_{pp}) + \Delta - \sqrt{ (2(t_{ss} + t_{pp})cos(ak) + \Delta)^2 + 16t_{sp}^2 \sin^2(ak)}.
\end{align}

\begin{figure}[H]
\includegraphics{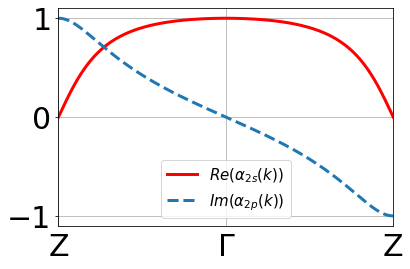}
\caption{Coefficients for the lower energy eigenstate obtained from the diagonalization of the tight binding Hamiltonian representing the 1D infinite Li chain}\label{fig:coeffs}
\end{figure}

While quantum states are only defined up to an overall phase, the phase relationship between these two coefficients is meaningful. We can see that the way we have currently presented them, that over the entire Brillouin zone, $\alpha_{2s}(k)$ is purely real and positive, while $\alpha_{2p}(k)$ is purely imaginary but is positive for $-\frac{\pi}{a} < k < 0$ and negative for $0 < k < \frac{\pi}{a}$. This is shown in Figure \ref{fig:coeffs}. The importance of these characteristics is explained in the main text.

%\bibliographystyle{apsrev}
%\bibliography{Susc_paper}

\end{document}